\documentclass[12pt,preprint]{aastex}

\usepackage{amssymb,graphicx,natbib}

\usepackage{graphics,color}
\usepackage{natbib}
\definecolor{dark}{gray}{0.5}
\definecolor{red}{rgb}{1,0,0}
\definecolor{green}{rgb}{0,1,0}
\definecolor{blue}{rgb}{0,0,1}
\def\n@space{\nulldelimiterspace=0pt\mathsurround=0pt}
\def\big#1{{\hbox{$\left#1\vbox to 8.5pt{}\right.\n@space$}}}
\def\Big#1{{\hbox{$\left#1\vbox to 11.5pt{}\right.\n@space$}}}
\def\bigg#1{{\hbox{$\left#1\vbox to 14.5pt{}\right.\n@space$}}}
\def\Bigg#1{{\hbox{$\left#1\vbox to 17.5pt{}\right.\n@space$}}}
\def\biggg#1{{\hbox{$\left#1\vbox to 20.5pt{}\right.\n@space$}}}
\def\Biggg#1{{\hbox{$\left#1\vbox to 23.5pt{}\right.\n@space$}}}

\def\dd{{\rm d}}
\def\half{{\textstyle{1\over2}}}
\def\threehalf{{\textstyle{3\over2}}}

\slugcomment{deformed spherical systems in MOND}

\shorttitle{deformed spherical systems in MOND}
\shortauthors{Chung-Ming Ko}

\begin{document}


\title{On the problem of deformed spherical systems in Modified Newtonian Dynamics}


\author{Chung-Ming Ko}
\affil{Institute of Astronomy, Department of Physics and Center for Complex Systems,\\
National Central University, Jhongli District, Taoyuan City, Taiwan 320, R.O.C.}
\email{cmko@astro.ncu.edu.tw}





\begin{abstract}

Based on Newtonian dynamics, observations show that the luminous masses of astrophysical objects that are the size of a galaxy or larger
are not enough to generate the measured motions
which they supposedly determine.
This is typically attributed to the existence of dark matter, which possesses mass but does not radiate (or absorb radiation).
Alternatively, the mismatch can be explained if the underlying dynamics is not Newtonian.
Within this conceptual scheme, Modified Newtonian Dynamics (MOND) is a successful theoretical paradigm.
MOND is usually expressed in terms of a nonlinear Poisson equation, which is difficult to analyse for arbitrary matter distributions.
We study the MONDian gravitational field generated by slightly non-spherically symmetric mass distributions
based on the fact that both Newtonian and MONDian fields are conservative (which we refer to as the compatibility condition).
As the non-relativistic version of MOND has two different formulations
(AQUAL and QuMOND) and the compatibility condition can be expressed in two ways,
there are four approaches to the problem in total.
The method involves solving a suitably defined linear deformation potential,
which generally depends on the choice of MOND interpolation function.
However, for some specific form of the deformation potential, the solution is independent of the interpolation function.

\end{abstract}


\section{Introduction}\label{sec:introduction}

The mass of an astrophysical object can be estimated using two methods.
The first one relies on the observed total amount of radiation emitted by the matter of the object
and the mass measured is called the luminous mass.
The second one depends on the motions of ambient objects (other objects or the object under investigation)
that are affected by the matter of the object and the mass measured is called the dynamical mass.
The first method requires a relation between the mass and the luminosity of the matter (either theoretical or empirical).
The relation is known as the ``mass-to-light ratio''.
The second method assumes that we understand the dynamical law governing the motions of the objects.
Newtonian dynamics (Newton's laws of motion supplemented by Newton's law of gravity) is well tested locally.
If we apply Newtonian dynamics to astrophysical objects that are the size of a galaxy or larger,
we generally find that the luminous mass is smaller than (usually much smaller than) the dynamical mass.
This mismatch in mass is usually called the ``missing mass problem''.
A logical solution, at this scale, is that the matter is dominated by a type of matter that possesses mass (thus provides gravity)
but does not emit or absorb electromagnetic radiation.
This type of matter is commonly known as dark matter.
We note that dark matter is also required in cosmology.
For a review of the history of dark matter, the reader is referred to the book by \citet{Sanders2010a}.

However, the mismatch in mass can be (and should be) interpreted in terms of a mismatch in acceleration:
the observed motion does not match the expected motion produced by the measured luminous mass if Newtonian dynamics is adopted.
What if Newtonian dynamics is not correct? This will open up explanations other than dark matter for the mismatch in acceleration
(or the ``missing mass problem'').
\citet{Milgrom1983a} proposed that when the acceleration is small
with respect to a characteristic scale (which is usually called the acceleration constant $a_0$),
Newton's second law of motion must be modified in order to
explain the mismatch (the acceleration must be larger than that predicted by Newton's law).
In subsequent papers, Milgrom provided a natural explanation to the flat rotation curve and Tully-Fisher relation of spiral galaxies,
the mass-to-light ratio of galaxy systems, etc. \citep[][]{Milgrom1983b,Milgrom1983c}.
This explanation was the birth of Modified Newtonian Dynamics (MOND).
We note that it is the scale of acceleration that distinguishes MOND from Newtonian dynamics, not other scales such as size, etc.
In the following year, \citet{Bekenstein1984} put the theory in a Lagrangian formulation which can be viewed as a
modified theory of gravity. Their theory is called Aquadratic Lagrangian theory (AQUAL).
\citet{Milgrom2010a} put forward another formulation of MOND called Quasi-linear formulation of MOND (QuMOND).
We will discuss in detail the two formulations in Section~\ref{sec:MONDformulations}.
MOND has been very successful in explaining many ``missing mass problems'' in galaxy-scale objects,
such as the flat rotation curve of spiral galaxies
\citep[e.g.,][]{Begeman1991,Sanders1996,deBlok1998,Sanders1998,Sanders2002,Famaey2005,Milgrom2007,Sanders2007,Swaters2010},
the baryonic Tully-Fisher relation \citep[e.g.,][]{McGaugh2005,McGaugh2011,McGaugh2012},
velocity dispersion in elliptical galaxies \citep[e.g.,][]{Milgrom2003,Chae2015,Tian2015},
the Faber-Jackson relation \citep[e.g.,][]{Sanders2010b},
and hot gas in elliptical galaxies \citep[e.g.,][]{Milgrom2012a}.
For the scales of cluster of galaxies, MOND is not as satisfactory. It seems that some form of dark matter is needed
\citep[see, e.g.,][]{Aguirre2001,Sanders2003,Clowe2003,
Angus2008}.
It is interesting to note that the gravitational redshift in galaxy clusters has also been studied in MOND
\citep[][]{Wojtak2011,Bekenstein2012}.

\citet{Bekenstein2004} proposed a covariant relativistic gravity theory called Tensor-Vector-Scalar theory (TeVeS) in which MOND
is the non-relativistic limit.
Later, \citet{Milgrom2009a} suggested another relativistic theory for MOND called BiMOND.
With a viable relativistic version of MOND, one can study relativistic phenomena such as gravitational lensing
\citep[see, e.g.,][]{Chiu2006,Zhao2006,Chiu2011,Milgrom2013,Tian2013,Sanders2014}
and cosmology \citep[see, e.g.,][]{Skordis2006,Skordisetal2006,Dodelson2006,Bourliot2007,Skordis2008,Skordis2009,Angus2009,Clifton2010,Milgrom2010b}.
\citet{Famaey2013} and \citet{McGaugh2015} pointed out challenges to both concordance $\Lambda$CDM cosmology and that of relativistic MOND.
We note that there are some theoretical issues to be sorted out in some forms of relativistic MOND theory
\citep[see, e.g.,][]{Seifert2007,Contaldi2008,Famaey2012}.

Although MOND was ``invented'' to study systems in the small acceleration regime, a number of studies have been devoted to the high acceleration regime
(i.e., regime close to the Newtonian limit), in particular, to the motion of objects in the Solar System,
such as the Pioneer anomaly, perihelion precession, etc.
\citep[see, e.g.,][]{Milgrom1983a,Milgrom2009b,Milgrom2012b,Sanders2006,Sereno2006,Iorio2008,Iorio2009,Iorio2010a,Iorio2010b,Iorio2013,
Sokaliwska2010,Blanchet2011,Hees2014,Hees2016}.
Precise measurements in the Solar System would place constraints on MOND (at least in the high acceleration regime).
Discussions of MOND in Solar System often involve the so-called ``external field effect'' \citep[EFE;][]{Milgrom1983a,Bekenstein1984}.
There is an absolute acceleration scale in MOND (the acceleration constant $a_0$), and thus the internal dynamics may depend on
the external gravitational field even if it is a uniform field (this violates the strong equivalence principle).
There are three characteristic accelerations: the gravitational acceleration by the internal field and that of the external field, and the
acceleration constant $a_0$ (hereafter gravitational field and acceleration will be used interchangeably).
Roughly speaking, if either the internal field or the external field is larger than $a_0$, then the internal dynamics will be governed by
standard Newtonian dynamics.
For those cases where $a_0$ is the largest, one finds the following:
(i) if the external field is larger than the internal field, then the internal dynamics will be Newtonian
but with a larger ``effective'' Newtonian gravitational constant;
(ii) if the internal field is larger than the external field, then the internal dynamics will be governed by MOND.
EFE is also important in the study of the dynamics of star clusters and satellite galaxies
\citep[see, e.g.,][]{Brada2000b,Baumgardt2005,Gentile2007,Haghi2009,Klypin2009,
Haghi2011,McGaugh2013a,McGaugh2013b,Derakhshani2014,Lughausen2014}.

It is worth noting that tabletop experiments on gravitational redshift using atom interferometers \citep[][]{Muller2010,Hohensee2011}
may be able to place some constraint on MOND in the high acceleration regime (see Appendix~\ref{sec:redshift}).

For more details on classical MOND, relativistic MOND, and other topics related to MOND,
the reader is referred to the excellent review by \citet{Famaey2012} and references therein.

As a modified theory of gravity, MOND can be expressed in terms of a nonlinear Poisson equation \citep[e.g.,][]{Bekenstein1984}
in which the Newtonian gravitational field and the MONDian field are related.
In general, the two fields differ by the curl of a vector, i.e., a solenoidal field, which in general depends on the
matter or mass distribution of the system (more on this in Section~\ref{sec:MONDformulations}).
As mentioned in \citet{Bekenstein1984}, the solenoidal field vanishes identically only if the system under investigation is highly symmetric
(e.g., planar, cylindrical, spherical).
For other systems, this term makes the analysis difficult and interesting.
Over the years, numerical schemes or solvers have been developed to solve the Poisson equation of less symmetric systems
\citep[see, e.g.,][]{Brada1999,Ciotti2006,Nipoti2007a,Tiret2007,Feix2008,Llinares2008,Londrillo2009,Angus2012,Candlish2015,Lughausen2015}.
These codes enable us to study, in the framework of MOND, the structure and evolution of stellar systems (mostly accompanied by an N-Body code),
such as stellar dynamics \citep[see,e.g.,][]{Nipoti2008,Nipoti2011},
disk galaxies \citep[see, e.g.,][]{Brada1999,Brada2000a,Tiret2007,Tiret2008,Angus2012,Lughausen2015},
elliptical galaxies \citep[see, e.g.,][]{Ciotti2006,Nipoti2007a,Wu2009,Wang2008,Wu2009},
satellite galaxies with an external field effect \citep[see, e.g.,][]{Brada2000b,Wu2007,Nipoti2007b,Haghi2011,Angus2014,Lughausen2014,Candlish2015},
gravitational lensing \citep[see, e.g.,][]{Feix2008},
and cosmic structure formation \citep[see, e.g.,][]{Llinares2008}.

Although the nonlinear Poisson equation is difficult to analyse analytically,
some progress has been made on
disk-like structures \citep[see, e.g.,][]{Brada1995}
and asymmetric or triaxial structures \citep[see, e.g.,][]{Angus2006,Ciotti2006,Shan2008,Ciotti2012}.
Analytic solutions have their role in our understanding of the systems and they are useful for testing numerical schemes.
This article explores analytically approximated solutions to slightly deformed spherical systems
\citep[cf. e.g.,][]{Milgrom1986,Ciotti2006}.

Both Newtonian and MONDian fields are conservative fields (i.e., expressible in terms of the gradient of a potential).
Both of their curls are identically zero.
We called the simultaneous curl-free requirement on both fields the compatibility condition.
Making use of this compatibility condition, we put forward an approximation scheme to solve the MONDian gravitational potential.
As there are two formulations of MOND (AQUAL and QuMOND) and the compatibility condition can be written in two ways,
we have four approaches to the problem altogether.
The paper is organized as follows.
Section~\ref{sec:MONDformulations} describes the two common formulations of MOND, AQUAL and QuMOND, and
their corresponding compatibility conditions.
Starting from a spherical system, we present treatments for slightly deformed systems for AQUAL and QuMOND
in Sections~\ref{sec:AQUALtreatment} and \ref{sec:QuMONDtreatment}, respectively.
A simple example is given in Section~\ref{sec:example} for illustration.
Section~\ref{sec:discussion} provides some discussions and remarks.

\section{Two formulations of MOND}\label{sec:MONDformulations}

MOND was invented as a modified law of inertia \citep{Milgrom1983a}.
Later, it was noticed that MOND can be (and is better) interpreted as a theory of modified gravity \citep[e.g.,][]{Bekenstein1984}.
Below, we present two formulations of MOND that were developed over the years: Aquadratic Lagrangian theory \citep[AQUAL,][]{Bekenstein1984}
and Quasi-linear formulation of MOND \citep[QuMOND,][]{Milgrom2010a}.

\subsection{AQUAL}\label{sec:AQUALformulation}

In AQUAL formulation, the gravitational acceleration in MOND is ${\bf g}_{\rm A}=-\nabla\Phi_{\rm A}$,
where the potential is given by the nonlinear Poisson equation
\begin{equation}\label{eq:AQUALpotential}
  \nabla\cdot\left[\tilde{\mu}(x_{\rm A})\nabla\Phi_{\rm A}\right]=4\pi G \rho=\nabla^2\Phi_{\rm N}\,,
  \quad
  x_{\rm A}={|\nabla\Phi_{\rm A}|\over a_0}={|{\bf g}_{\rm A}|\over a_0}\,,
\end{equation}
where $\Phi_{\rm N}$ is the Newtonian gravitational potential.
Here, $a_0$ is the utmost important acceleration constant of MOND.
$\tilde{\mu}(x_{\rm A})$ is called the interpolation function in AQUAL, and
$\tilde{\mu}(x_{\rm A})\rightarrow 1$ as $x_{\rm A}\rightarrow \infty$,
and $\tilde{\mu}(x_{\rm A})\rightarrow x_{\rm A}$ as $x\rightarrow 0$
(i.e., Newtonian regime and deep MOND regime, respectively).
Different forms of the interpolation function have been used in the literature.
The most commonly used forms are, e.g., the standard form proposed by \citet{Milgrom1983a},
\begin{equation}\label{eq:standardform}
  \tilde{\mu}(x_{\rm A})={x_{\rm A}\over\sqrt{1+x_{\rm A}^2\,}}\,,
\end{equation}
the simple form by \citet{Famaey2005},
\begin{equation}\label{eq:simpleform}
  \tilde{\mu}(x_{\rm A})={x_{\rm A}\over(1+x_{\rm A})}\,,
\end{equation}
and the Bekenstein form by \citet{Bekenstein2004},
\begin{equation}\label{eq:Bekensteinform}
  \tilde{\mu}(x_{\rm A})={-1+\sqrt{1+4x_{\rm A}\,}\over 1+\sqrt{1+4x_{\rm A}\,}}\,.
\end{equation}
All of these forms (and some others in the literature) can be included in the two-parameter
canonical form proposed by \citet{Chiu2011}:
\begin{equation}\label{eq:canonicalmu}
  \tilde{\mu}(x_{\rm A})=\left[1-{2\over (1+\eta x_{\rm A}^\alpha)+\sqrt{(1-\eta x_{\rm A}^\alpha)^2+4x_{\rm A}^\alpha\,}}\right]^{1/\alpha}\,,
\end{equation}
where $\alpha>0$ and $\eta\ge 0$.
Here, $(\alpha,\eta)=(1,0)$, $(1,1)$, and $(2,1)$ correspond to the Bekenstein form, the simple form, and the standard form, respectively.

Integrating Equation~(\ref{eq:AQUALpotential}) once gives
\begin{equation}\label{eq:AQUALacc}
  \tilde{\mu}(x_{\rm A}){\bf g}_{\rm A}=-\tilde{\mu}(x_{\rm A})\nabla\Phi_{\rm A}=-\nabla\Phi_{\rm N}+\nabla\times{\bf h}
  ={\bf g}_{\rm N}+\nabla\times{\bf h}={\bf G}_{\rm N}\,,
\end{equation}
where ${\bf g}_{\rm N}=-\nabla\Phi_{\rm N}$ is the Newtonian gravitational acceleration.
Here, ${\bf h}$ is an arbitrary vector.
Inverting Equation~(\ref{eq:AQUALacc}) gives
\begin{equation}\label{eq:AQUALinvert}
  -\nabla\Phi_{\rm A}={\bf g}_{\rm A}=\tilde{\nu}(\chi_{\rm N}){\bf G}_{\rm N}
  =\tilde{\nu}(\chi_{\rm N})\left({\bf g}_{\rm N}+\nabla\times{\bf h}\right)
  =\tilde{\nu}(\chi_{\rm N})\left(-\nabla\Phi_{\rm N}+\nabla\times{\bf h}\right)\,,
\end{equation}
where
\begin{equation}\label{eq:AQUALchi}
  \chi_{\rm N}={|{\bf G}_{\rm N}|\over a_0}={1\over a_0}\left|{\bf g}_{\rm N}+\nabla\times{\bf h}\right|
  ={1\over a_0}\left|-\nabla\Phi_{\rm N}+\nabla\times{\bf h}\right|\,.
\end{equation}
$\tilde{\nu}(\chi_{\rm N})$ is called the inverted interpolation function in AQUAL.
The inverted form corresponding to the canonical form of Equation~(\ref{eq:canonicalmu}) is \citep{Chiu2011}
\begin{equation}\label{eq:canonicalnu}
  \tilde{\nu}(\chi_{\rm N})=\left[1+\half\left(\sqrt{4\chi_{\rm N}^{-\alpha}+\eta^2}-\eta\right)\right]^{1/\alpha}\,.
\end{equation}

Since ${\bf g}_{\rm A}$ and ${\bf g}_{\rm N}$ can be expressed as a gradient of a potential,
there exists a compatibility condition for Equations~(\ref{eq:AQUALacc}) and (\ref{eq:AQUALinvert}).
Taking the curl of Equation~(\ref{eq:AQUALacc}) gives
\begin{equation}\label{eq:AQUALcompatibleN}
  0=\nabla\times{\bf g}_{\rm N}=\nabla\times\left(\tilde{\mu}{\bf g}_{\rm A}\right)-\nabla\times\nabla\times{\bf h}
  ={1\over a_0}{\dd\tilde{\mu}\over\dd x_{\rm A}}\left(\nabla|{\bf g}_{\rm A}|\right)\times{\bf g}_{\rm A}
  -\nabla\times\nabla\times{\bf h}\,,
\end{equation}
and taking the curl of Equation~(\ref{eq:AQUALinvert}) gives
\begin{equation}\label{eq:AQUALcompatibleA}
  0=\nabla\times{\bf g}_{\rm A}=\nabla\times\left(\tilde{\nu}{\bf G}_{\rm N}\right)
  ={1\over a_0}{\dd\tilde{\nu}\over\dd\chi_{\rm N}}\left(\nabla|{\bf G}_{\rm N}|\right)\times{\bf G}_{\rm N}
  +\tilde{\nu}\nabla\times{\bf G}_{\rm N}\,.
\end{equation}

For highly symmetric systems (such as, planar, cylindrical, spherical), $\nabla\times{\bf h}=0$.
Consequently, $\chi_{\rm N}=x_{\rm N}=|{\bf g}_{\rm N}|/a_0$,
\begin{equation}\label{eq:AQUALsym}
  \tilde{\mu}(x_{\rm A}){\bf g}_{\rm A}={\bf g}_{\rm N}\,,
  \quad
  {\bf g}_{\rm A}=\tilde{\nu}(x_{\rm N}){\bf g}_{\rm N}\,.
\end{equation}
and the compatibility conditions, Equations~(\ref{eq:AQUALcompatibleN}) and (\ref{eq:AQUALcompatibleA}), are satisfied automatically.

\subsection{QuMOND}\label{sec:QuMONDformulation}

In QuMOND formulation, the gravitational acceleration in MOND is ${\bf g}_{\rm Q}=-\nabla\Phi_{\rm Q}$,
where the potential is given by
\begin{equation}\label{eq:QuMONDpotential}
  \nabla^2\Phi_{\rm Q}=\nabla\cdot\left[\nu(x_{\rm N})\nabla\Phi_{\rm N}\right]\,,
  \quad
  \nabla^2\Phi_{\rm N}=4\pi G\rho\,,
  \quad
  x_{\rm N}={|\nabla\Phi_{\rm N}|\over a_0}={|{\bf g}_{\rm N}|\over a_0}\,.
\end{equation}
$\nu(x_{\rm N})$ is called the inverted interpolation function in QuMOND,
and $\nu(x_{\rm N})\rightarrow 1$ as $x_{\rm N}\rightarrow \infty$,
and $\nu(x_{\rm N})\rightarrow 1/\sqrt{x_{\rm N}}$ as $x_{\rm N}\rightarrow 0$
(i.e., Newtonian regime and deep MOND regime, respectively).
A useful form for the inverted interpolation function is Equation~(\ref{eq:canonicalnu}) (with $\chi_{\rm N}$ replaced by $x_{\rm N}$).

Integrating Equation~(\ref{eq:QuMONDpotential}) once gives
\begin{equation}\label{eq:QuMONDinvert}
  \nu(x_{\rm N}){\bf g}_{\rm N}=-\nu(x_{\rm N})\nabla\Phi_{\rm N}=-\nabla\Phi_{\rm Q}-\nabla\times{\bf A}
  ={\bf g}_{\rm Q}-\nabla\times{\bf A}={\bf G}_{\rm Q}\,.
\end{equation}
Here, ${\bf A}$ is an arbitrary vector.
Inverting Equation~(\ref{eq:QuMONDinvert}) gives (cf. Equation~(\ref{eq:AQUALacc}))
\begin{equation}\label{eq:QuMONDacc}
  -\nabla\Phi_{\rm N}={\bf g}_{\rm N}=\mu(\chi_{\rm Q}){\bf G}_{\rm Q}
  =\mu(\chi_{\rm Q})\left({\bf g}_{\rm Q}-\nabla\times{\bf A}\right)
  =\mu(\chi_{\rm Q})\left(-\nabla\Phi_{\rm Q}-\nabla\times{\bf A}\right)\,,
\end{equation}
where (cf. Equation~(\ref{eq:AQUALchi}))
\begin{equation}\label{eq:QuMONDchi}
  \chi_{\rm Q}={|{\bf G}_{\rm Q}|\over a_0}={1\over a_0}\left|{\bf g}_{\rm Q}-\nabla\times{\bf A}\right|
  ={1\over a_0}\left|-\nabla\Phi_{\rm Q}-\nabla\times{\bf A}\right|\,.
\end{equation}
$\mu(\chi_{\rm Q})$ is called the interpolation function in QuMOND.
A useful form for the interpolation function is Equation~(\ref{eq:canonicalmu}) (with $x_{\rm A}$ replaced by $\chi_{\rm Q}$).

Similar to AQUAL,
there exists a compatibility condition for Equations~(\ref{eq:QuMONDinvert}) and (\ref{eq:QuMONDacc}).
Taking the curl of Equation~(\ref{eq:QuMONDinvert}) gives (cf. Equation~(\ref{eq:AQUALcompatibleN}))
\begin{equation}\label{eq:QuMONDcompatibleQ}
  0=\nabla\times{\bf g}_{\rm Q}=\nabla\times\left(\nu{\bf g}_{\rm N}\right)+\nabla\times\nabla\times{\bf A}
  ={1\over a_0}{\dd\nu\over\dd x_{\rm N}}\left(\nabla|{\bf g}_{\rm N}|\right)\times{\bf g}_{\rm N}
  +\nabla\times\nabla\times{\bf A}\,,
\end{equation}
and taking the curl of Equation~(\ref{eq:QuMONDacc}) gives (cf. Equation~(\ref{eq:AQUALcompatibleA}))
\begin{equation}\label{eq:QuMONDcompatibleN}
  0=\nabla\times{\bf g}_{\rm N}=\nabla\times\left(\mu{\bf G}_{\rm Q}\right)
  ={1\over a_0}{\dd\mu\over\dd\chi_{\rm Q}}\left(\nabla|{\bf G}_{\rm Q}|\right)\times{\bf G}_{\rm Q}
  +\mu\nabla\times{\bf G}_{\rm Q}\,.
\end{equation}

Similar to AQUAL, for highly symmetric systems (such as, planar, cylindrical, spherical), $\nabla\times{\bf A}=0$.
Consequently, $\chi_{\rm Q}=x_{\rm Q}=|{\bf g}_{\rm Q}|/a_0$,
\begin{equation}\label{eq:QuMONDsym}
  \nu(x_{\rm N}){\bf g}_{\rm N}={\bf g}_{\rm Q}\,,
  \quad
  {\bf g}_{\rm N}=\mu(x_{\rm Q}){\bf g}_{\rm Q}\,.
\end{equation}
and the compatibility conditions, Equations~(\ref{eq:QuMONDcompatibleQ}) and (\ref{eq:QuMONDcompatibleN}), are satisfied automatically.

\section{Systems slightly deformed from spherical symmetry}\label{sec:deform}

To find the MONDian gravitational acceleration produced by a general mass distribution $\rho$ is a formidable task.
One has to solve $\nabla\cdot\left[\tilde{\mu}(|{\bf g}_{\rm A}|/a_0){\bf g}_{\rm A}\right]=\nabla\cdot{\bf g}_{\rm N}$ in AQUAL formulation
or $\nabla\cdot{\bf g}_{\rm Q}=\nabla\cdot\left[\nu(|{\bf g}_{\rm N}|/a_0){\bf g}_{\rm N}\right]$ in QuMOND formulation, with
the Newtonian gravitational acceleration given by $\nabla\cdot{\bf g}_{\rm N}=-4\pi G\rho$.
However, for a spherical mass distribution $\rho(r)$, the solution can be written down as follows.
The Newtonian gravitational acceleration is given by
\begin{equation}\label{eq:NewtonSpherical}
  {\bf g}_{\rm N}=-\,{G m(r)\over r^2}\,{\hat{\bf e}}_r\,,
  \quad
  m(r)=\int_0^r 4\pi \rho(r^\prime)\dd r^\prime\,.
\end{equation}
Here, $m(r)$ is the mass within radius $r$.
The MONDian gravitational acceleration is given by
${\bf g}_{\rm A}=\tilde{\nu}(|{\bf g}_{\rm N}/a_0|){\bf g}_{\rm N}$ in AQUAL formulation and
${\bf g}_{\rm Q}=\nu(|{\bf g}_{\rm N}/a_0|){\bf g}_{\rm N}$ in QuMOND formulation
(provided that the interpolation function $\tilde\nu$ and $\nu$ are known).
The compatibility conditions mentioned in Section~\ref{sec:MONDformulations} are satisfied automatically.

Base on the spherical solution, we propose a treatment for slightly deformed spherical systems.
Our goal is to find solutions that will at least approximately satisfy the compatibility condition.
Since we can express the compatibility condition in terms of the interpolation function ($\tilde{\mu}$ in AQUAL or $\mu$ in QuMOND)
or its inverse ($\tilde{\nu}$ in AQUAL or $\nu$ in QuMOND), we have four schemes:
Equation~(\ref{eq:AQUALcompatibleA}) and (\ref{eq:AQUALcompatibleN}) for AQUAL
and Equations~(\ref{eq:QuMONDcompatibleQ}) and (\ref{eq:QuMONDcompatibleN}) for QuMOND, respectively.
We present the four schemes in detail in the following.
We note that on the one hand, AQUAL and MOND are not exactly equivalent (except for the spherical, cylindrical, planar cases).
On the other hand, choosing an interpolation function or its inverse is a matter of convenience.

\subsection{Treatment in AQUAL}\label{sec:AQUALtreatment}

There are two expressions for the compatibility condition.
We thus have two schemes: AQUAL I based on
$\nabla\times{\bf g}_{\rm A}=\nabla\times\left[\tilde{\nu}\left({\bf g}_{\rm N}+\nabla\times{\bf h}\right)\right]=0$
(Equation~(\ref{eq:AQUALinvert})), and
AQUAL II based on
$\nabla\times{\bf g}_{\rm N}=\nabla\times\left(\tilde{\mu}{\bf g}_{\rm A}-\nabla\times{\bf h}\right)=0$
(Equation~(\ref{eq:AQUALacc})).

\subsubsection{AQUAL I: $\nabla\times{\bf g}_{\rm A}=\nabla\times\left[\tilde{\nu}\left({\bf g}_{\rm N}+\nabla\times{\bf h}\right)\right]=0$}
\label{sec:AQUALI}

Suppose the Newtonian gravitational acceleration deviates slightly from spherical symmetry:
\begin{equation}\label{eq:AQUALIgN}
  {\bf g}_{\rm N}=g_{\rm N}^{(0)}\,{\hat{\bf e}}_r+\epsilon\, {\bf g}_{\rm N}^{(1)}+{\cal O}(\epsilon^2)\,,
\end{equation}
where $\epsilon$ is the small parameter keeping track of the order, $g_{\rm N}^{(0)}=g_{\rm N}^{(0)}(r)$ depends on $r$ only, and
${\bf g}_{\rm N}^{(1)}={\bf g}_{\rm N}^{(1)}(r,\theta,\phi)$ depends on $(r,\theta,\phi)$.
Since for a spherically symmetric system $\nabla\times{\bf h}=0$, we expect $\nabla\times{\bf h}={\cal O}(\epsilon)$ in slightly
deformed spherical systems (i.e., non-spherical), and we replace it by $\epsilon\nabla\times{\bf h}_{\rm N}^{(1)}$.
Thus, Equation~(\ref{eq:AQUALacc}) becomes
\begin{equation}\label{eq:AQUALIGN}
  {\bf G}_{\rm N}=g_{\rm N}^{(0)}\,{\hat{\bf e}}_r+\epsilon\,\left[ {\bf g}_{\rm N}^{(1)}+\nabla\times{\bf h}_{\rm N}^{(1)}\right]+{\cal O}(\epsilon^2)
  =g_{\rm N}^{(0)}\,{\hat{\bf e}}_r+\epsilon\,{\bf F}_{\rm N}^{(1)}+{\cal O}(\epsilon^2)\,.
\end{equation}
Therefore, up to ${\cal O}(\epsilon)$, the compatibility condition Equation~(\ref{eq:AQUALcompatibleA}) becomes
\begin{equation}\label{eq:AQUALIcompatible}
  0={\tilde{\cal A}}^{(0)}\left[g_{\rm N}^{(0)}\nabla F_{{\rm N}r}^{(1)}
  -{\dd g_{\rm N}^{(0)}\over\dd r}{\bf F}_{\rm N}^{(1)}\right]\times{\hat{\bf e}}_r+\nabla\times{\bf F}_{\rm N}^{(1)}\,,
  \quad
  {\tilde{\cal A}}^{(0)}=\left[{1\over a_0\tilde{\nu}}{\dd\tilde{\nu}\over\dd\chi_{\rm N}}\right]^{(0)}\,.
\end{equation}
Here, $[\,\cdot\,]^{(0)}$ denotes quantities of ${\cal O}(1)$ (note $a_0\chi_{\rm N}=g_{\rm N}^{(0)}+{\cal O}(\epsilon)$),
i.e., evaluated at the spherically symmetric level. Hence, ${\tilde{\cal A}}^{(0)}={\tilde{\cal A}}^{(0)}(r)$ depends on $r$ only.
Solving Equation~(\ref{eq:AQUALIcompatible}) implies that ${\bf F}_{\rm N}^{(1)}$ can be expressed in terms of
a ``deformation potential'' $\Psi_{\rm N}^{(1)}(r,\theta,\phi)$,
\begin{equation}\label{eq:AQUALIFN}
  {\bf F}_{\rm N}^{(1)}=-\nabla\Psi^{(1)}_{\rm N}+f_{\rm N}^{(1)}\,{\hat{\bf e}}_r\,,
  \quad
  f_{\rm N}^{(1)}={{\tilde{\cal A}}^{(0)}\over[{\tilde{\cal A}}^{(0)}g_{\rm N}^{(0)}+1]}
  \left[g_{\rm N}^{(0)}{\partial\Psi^{(1)}_{\rm N}\over\partial r}
  -{\dd g_{\rm N}^{(0)}\over\dd r}\Psi^{(1)}_{\rm N}\right]\,.
\end{equation}
By the Helmholtz theorem, we can express
$f_{\rm N}^{(1)}\,{\hat{\bf e}}_r=-\nabla\varphi^{(1)}_{\rm N}+\nabla\times{\bf h}^{(1)}_{\rm N}$,
with
\begin{eqnarray}\label{eq:fNphih}
  \varphi^{(1)}_{\rm N}&=&
  {\displaystyle
  \int_V{\nabla^\prime\cdot[f_{\rm N}^{(1)}({\bf x}^\prime)\,{\hat{\bf e}}_{r^\prime}]
  \over 4\pi|{\bf x}-{\bf x}^\prime|}\,\dd^3 x^{\prime}
  -\,\oint_S {[f_{\rm N}^{(1)}({\bf x}^\prime)\,{\hat{\bf e}}_{r^\prime}]\cdot{\hat{\bf n}^\prime}
  \over 4\pi|{\bf x}-{\bf x}^\prime|}\,\dd^2 x^{\prime}\,,} \\
  {\bf h}^{(1)}_{\rm N}&=&
  {\displaystyle
  \int_V{\nabla^\prime\times[f_{\rm N}^{(1)}({\bf x}^\prime)\,{\hat{\bf e}}_{r^\prime}]
  \over 4\pi|{\bf x}-{\bf x}^\prime|}\,\dd^3 x^{\prime}
  +\,\oint_S {[f_{\rm N}^{(1)}({\bf x}^\prime)\,{\hat{\bf e}}_{r^\prime}]\times{\hat{\bf n}^\prime}
  \over 4\pi|{\bf x}-{\bf x}^\prime|}\,\dd^2 x^{\prime}\,.}
\end{eqnarray}
If the $f_{\rm N}^{(1)}$ decay rapid enough as $|{\bf x}^\prime|$ tends to infinity, then as the integral is extended to the
entire space, the surface terms will vanish.
The relation between ${\bf F}_{\rm N}^{(1)}$ and ${\bf g}_{\rm N}^{(1)}$ (Equation~(\ref{eq:AQUALIGN})) gives
\begin{equation}\label{eq:AQUALIgNresult}
  {\bf g}_{\rm N}=g_{\rm N}^{(0)}\,{\hat{\bf e}}_r-\epsilon\,\nabla\left[\Psi^{(1)}_{\rm N}+\varphi^{(1)}_{\rm N}\right]+{\cal O}(\epsilon^2)\,,
\end{equation}
Moreover, Equation~(\ref{eq:AQUALIGN}) gives
\begin{equation}\label{eq:AQUALIchiN}
  |{\bf G}_{\rm N}|=a_0\chi_{\rm N}=a_0\left[\chi_{\rm N}^{(0)}+\epsilon\,\chi_{\rm N}^{(1)}+{\cal O}(\epsilon^2)\right]
  =g_{\rm N}^{(0)}+\epsilon\,F_{{\rm N}r}^{(1)}+{\cal O}(\epsilon^2)\,,
\end{equation}
and
\begin{equation}\label{eq:AQUALInu}
  {\tilde\nu}(\chi_{\rm N})={\tilde\nu}^{(0)}
  +\epsilon\, {\tilde\nu}^{(0)}{\tilde{\cal A}}^{(0)}\,a_0\chi_{\rm N}^{(1)}+{\cal O}(\epsilon^2)
  ={\tilde\nu}^{(0)}+\epsilon\, {\tilde\nu}^{(0)}{\tilde{\cal A}}^{(0)}F_{{\rm N}r}^{(1)}+{\cal O}(\epsilon^2)\,,
\end{equation}
where ${\tilde\nu}^{(0)}={\tilde\nu}(\chi_{\rm N}^{(0)})$.
Hence,
\begin{equation}\label{eq:AQUALIgAresult}
  {\bf g}_{\rm A}={\tilde\nu}^{(0)}g_{\rm N}^{(0)}\,{\hat{\bf e}}_r-\epsilon\,{\tilde\nu}^{(0)}\left[\nabla\Psi^{(1)}_{\rm N}
  +{\tilde{\cal A}}^{(0)}{\dd g_{\rm N}^{(0)}\over\dd r}\Psi^{(1)}_{\rm N}\,{\hat{\bf e}}_r\right]+{\cal O}(\epsilon^2)\,.
\end{equation}
The mass distribution is given by the Poisson equation $4\pi G\rho=\nabla\cdot{\bf g}_{\rm N}$.
If we express $\rho=\rho^{(0)}(r)+\epsilon\,\rho^{(1)}(r,\theta,\phi)+{\cal O}(\epsilon^2)$ and make use of
Equations~(\ref{eq:AQUALIgNresult}) and (\ref{eq:AQUALIFN})
(and $f_{\rm N}^{(1)}\,{\hat{\bf e}}_r=-\nabla\varphi^{(1)}_{\rm N}+\nabla\times{\bf h}^{(1)}_{\rm N}$),
then we obtain
\begin{equation}\label{eq:AQUALIrho0}
  4\pi G\rho^{(0)}=-\,{1\over r^2}{\partial r^2 g_N^{(0)}\over\partial r}\,,
\end{equation}
and
\begin{equation}\label{eq:AQUALIrho1}
  4\pi G\rho^{(1)}=\nabla^2\Psi_{\rm N}^{(1)}-\,{1\over r^2}{\!\partial\over\partial r}\left\{
  {r^2{\tilde{\cal A}}^{(0)}\over[{\tilde{\cal A}}^{(0)}g_{\rm N}^{(0)}+1]}
  \left[g_{\rm N}^{(0)}{\partial\Psi^{(1)}_{\rm N}\over\partial r}-{\dd g_{\rm N}^{(0)}\over\dd r}\Psi^{(1)}_{\rm N}\right]\right\}\,.
\end{equation}

\subsubsection{AQUAL II: $\nabla\times{\bf g}_{\rm N}=\nabla\times\left(\tilde{\mu}{\bf g}_{\rm A}-\nabla\times{\bf h}\right)=0$}
\label{sec:AQUALII}

Suppose the MONDian gravitational acceleration deviates slightly from spherical symmetry:
\begin{equation}\label{eq:AQUALIIgA}
  {\bf g}_{\rm A}=g_{\rm A}^{(0)}\,{\hat{\bf e}}_r+\epsilon\, {\bf g}_{\rm A}^{(1)}+{\cal O}(\epsilon^2)
  =g_{\rm A}^{(0)}\,{\hat{\bf e}}_r-\epsilon\, \nabla\Phi_{\rm A}^{(1)}+{\cal O}(\epsilon^2)\,.
\end{equation}
Here, $g_{\rm A}^{(0)}=g_{\rm A}^{(0)}(r)$ depends on $r$ only,
${\bf g}_{\rm A}^{(1)}={\bf g}_{\rm A}^{(1)}(r,\theta,\phi)$ depends on $(r,\theta,\phi)$, and
$\Phi_{\rm A}^{(1)}((r,\theta,\phi)$ can be called the ``deformation potential''.
Besides, we expect $\nabla\times{\bf h}={\cal O}(\epsilon)$ in slightly non-spherical systems
because $\nabla\times{\bf h}=0$ for spherical systems.
We replace the curl term by $\epsilon\nabla\times{\bf h}_{\rm A}^{(1)}$.

Up to ${\cal O}(\epsilon)$, the compatibility condition (Equation~(\ref{eq:AQUALcompatibleN}))
becomes
\begin{equation}\label{eq:AQUALIIcompatible}
  0=\tilde{\mu}^{(0)}{\tilde{\cal B}}^{(0)}\left[g_{\rm A}^{(0)}\nabla g_{{\rm A}r}^{(1)}
  -{\dd g_{\rm A}^{(0)}\over\dd r}{\bf g}_{\rm A}^{(1)}\right]\times{\hat{\bf e}}_r
  -\nabla\times\nabla\times{\bf h}_{\rm A}^{(1)}\,,
  \quad
  {\tilde{\cal B}}^{(0)}=\left[{1\over a_0\tilde{\mu}}{\dd\tilde{\mu}\over\dd x_{\rm A}}\right]^{(0)}\,.
\end{equation}
Here, $[\,\cdot\,]^{(0)}$ denotes quantities of ${\cal O}(1)$ (note $a_0 x_{\rm A}=g_{\rm A}^{(0)}+{\cal O}(\epsilon)$),
i.e., evaluated at the spherically symmetric level. Hence, ${\tilde{\cal B}}^{(0)}={\tilde{\cal B}}^{(0)}(r)$ depends on $r$ only.
Solving Equation~(\ref{eq:AQUALIIcompatible}) with ${\bf g}_{\rm A}^{(1)}=-\nabla\Phi_{\rm A}^{(1)}$ implies that
\begin{equation}\label{eq:AQUALIIqA}
  q_{\rm A}^{(1)}\,{\hat{\bf e}}_r=-\nabla\Upsilon_{\rm A}^{(1)}
  -\nabla\times\nabla\times{\bf h}_{\rm A}^{(1)}\,,
  \quad
  q_{\rm A}^{(1)}=\tilde{\mu}^{(0)}{\tilde{\cal B}}^{(0)}
  \left[g_{\rm A}^{(0)}{\partial\Phi^{(1)}_{\rm A}\over\partial r}-{\dd g_{\rm A}^{(0)}\over\dd r}\Phi^{(1)}_{\rm A}\right]\,.
\end{equation}
The Helmholtz theorem gives
\begin{eqnarray}\label{eq:qAUpsilonh}
  \Upsilon^{(1)}_{\rm A}&=&
  {\displaystyle
  \int_V{\nabla^\prime\cdot[q_{\rm A}^{(1)}({\bf x}^\prime)\,{\hat{\bf e}}_{r^\prime}]
  \over 4\pi|{\bf x}-{\bf x}^\prime|}\,\dd^3 x^{\prime}
  -\,\oint_S {[q_{\rm A}^{(1)}({\bf x}^\prime)\,{\hat{\bf e}}_{r^\prime}]\cdot{\hat{\bf n}^\prime}
  \over 4\pi|{\bf x}-{\bf x}^\prime|}\,\dd^2 x^{\prime}\,,} \\
  {\bf h}^{(1)}_{\rm A}&=&
  {\displaystyle
  -\int_V{\nabla^\prime\times[q_{\rm A}^{(1)}({\bf x}^\prime)\,{\hat{\bf e}}_{r^\prime}]
  \over 4\pi|{\bf x}-{\bf x}^\prime|}\,\dd^3 x^{\prime}
  -\,\oint_S {[q_{\rm A}^{(1)}({\bf x}^\prime)\,{\hat{\bf e}}_{r^\prime}]\times{\hat{\bf n}^\prime}
  \over 4\pi|{\bf x}-{\bf x}^\prime|}\,\dd^2 x^{\prime}\,.}
\end{eqnarray}
If $q_{\rm A}^{(1)}$ decay rapidly enough as $|{\bf x}^\prime|$ tends to infinity, then as the integral is extended to the
entire space, the surface terms will vanish.

Note that
\begin{equation}\label{eq:AQUALIImu}
  \tilde{\mu}(x_{\rm A}){\bf g}_{\rm A}=\tilde{\mu}^{(0)}g_{\rm A}^{(0)}\,{\hat{\bf e}}_r -\epsilon\, \tilde{\mu}^{(0)}
  \left[\nabla\Phi_{\rm A}^{(1)}+{\tilde{\cal B}}^{(0)}g_{\rm A}^{(0)}{\partial\Phi_{\rm A}^{(1)}\over\partial r}\,{\hat{\bf e}}_r\right]
  +{\cal O}(\epsilon^2)\,,
\end{equation}
and thus
\begin{eqnarray}\label{eq:AQUALIIgNresult}
  {\bf g}_{\rm N}&=&\tilde{\mu}^{(0)}g_{\rm A}^{(0)}\,{\hat{\bf e}}_r +\epsilon\, \left\{\tilde{\mu}^{(0)}
  \left[{\bf g}_{\rm A}^{(1)}+{\tilde{\cal B}}^{(0)}g_{\rm A}^{(0)}g_{{\rm A}r}^{(1)}\,{\hat{\bf e}}_r\right]
  -\nabla\times{\bf h}_{\rm A}^{(1)}\right\}+{\cal O}(\epsilon^2) \nonumber \\
  &=&\tilde{\mu}^{(0)}g_{\rm A}^{(0)}\,{\hat{\bf e}}_r-\epsilon\,\left\{\tilde{\mu}^{(0)}\left[\nabla\Phi^{(1)}_{\rm A}
  +{\tilde{\cal B}}^{(0)}{\dd g_{\rm A}^{(0)}\over\dd r}\Phi^{(1)}_{\rm A}\,{\hat{\bf e}}_r\right]-\nabla\Upsilon_{\rm A}^{(1)}\right\}
  +{\cal O}(\epsilon^2)\,.
\end{eqnarray}
Equation~(\ref{eq:AQUALIImu}) or (\ref{eq:AQUALIIgNresult}) gives the mass distribution
\begin{equation}\label{eq:AQUALIIrho0}
  4\pi G\rho^{(0)}=-\,{1\over r^2}{\partial r^2 g_N^{(0)}\over\partial r}
  =-\,{1\over r^2}{\partial r^2 \tilde{\mu}^{(0)}g_{\rm A}^{(0)}\over\partial r}\,,
\end{equation}
and
\begin{equation}\label{eq:AQUALIIrho1}
  4\pi G \rho^{(1)}=\nabla\cdot\left\{\tilde{\mu}^{(0)}\left[\nabla\Phi^{(1)}_{\rm A}
  +{\tilde{\cal B}}^{(0)}g_{\rm A}^{(0)}{\dd\Phi^{(1)}_{\rm A}\over\dd r}\,{\hat{\bf e}}_r\right]\right\}\,.
\end{equation}

\subsection{Treatment in QuMOND}\label{sec:QuMONDtreatment}

Similar to AQUAL, there are two expressions for the compatibility
condition, and hence we also have two schemes: QuMOND I based on
$\nabla\times{\bf g}_{\rm N}=\nabla\times\left[\mu\left({\bf g}_{\rm Q}-\nabla\times{\bf A}\right)\right]=0$
(Equation~(\ref{eq:QuMONDacc})) and QuMOND II based on
$\nabla\times{\bf g}_{\rm Q}=\nabla\times\left(\nu{\bf g}_{\rm N}+\nabla\times{\bf A}\right)=0$ (Equation~(\ref{eq:QuMONDinvert})).

\subsubsection{QuMOND I: $\nabla\times{\bf g}_{\rm N}=\nabla\times\left[\mu\left({\bf g}_{\rm Q}-\nabla\times{\bf A}\right)\right]=0$}
\label{sec:QuMONDI}

The mathematical procedure is the same as in Section~\ref{sec:AQUALI}.
All we need to do is to change ${\bf g}_{\rm N}$ to ${\bf g}_{\rm Q}$, ${\bf h}$ to $-{\bf A}$, ${\tilde{\nu}}$ to $\mu$, etc.
We start from
\begin{equation}\label{eq:QuMONDIgQ}
  {\bf g}_{\rm Q}=g_{\rm Q}^{(0)}\,{\hat{\bf e}}_r+\epsilon\, {\bf g}_{\rm Q}^{(1)}+{\cal O}(\epsilon^2)\,,
\end{equation}
and get
\begin{equation}\label{eq:QuMONDIGQ}
  {\bf G}_{\rm Q}=g_{\rm Q}^{(0)}\,{\hat{\bf e}}_r+\epsilon\,\left[ {\bf g}_{\rm Q}^{(1)}-\nabla\times{\bf A}_{\rm Q}^{(1)}\right]+{\cal O}(\epsilon^2)
  =g_{\rm Q}^{(0)}\,{\hat{\bf e}}_r+\epsilon\,{\bf F}_{\rm Q}^{(1)}+{\cal O}(\epsilon^2)\,.
\end{equation}
The compatibility condition Equation~(\ref{eq:QuMONDcompatibleN}) becomes
\begin{equation}\label{eq:QuMONDIcompatible}
  0={\cal B}^{(0)}\left[g_{\rm Q}^{(0)}\nabla F_{{\rm Q}r}^{(1)}
  -{\dd g_{\rm Q}^{(0)}\over\dd r}{\bf F}_{\rm Q}^{(1)}\right]\times{\hat{\bf e}}_r
  +\nabla\times{\bf F}_{\rm Q}^{(1)}\,,
  \quad
  {\cal B}^{(0)}=\left[{1\over a_0\mu}{\dd\mu\over\dd\chi_{\rm Q}}\right]^{(0)}\,.
\end{equation}
Solving Equation~(\ref{eq:QuMONDIcompatible}) implies that ${\bf F}_{\rm Q}^{(1)}$ can be expressed in terms of
a ``deformation potential'' $\Psi_{\rm Q}^{(1)}(r,\theta,\phi)$,
\begin{equation}\label{eq:QuMONDIFQ}
  {\bf F}_{\rm Q}^{(1)}=-\nabla\Psi_{\rm Q}^{(1)}+f_{\rm Q}^{(1)}\,{\hat{\bf e}}_r\,,
  \quad
  f_{\rm Q}^{(1)}={{\cal B}^{(0)}\over[{\cal B}^{(0)}g_{\rm Q}^{(0)}+1]}
  \left[g_{\rm Q}^{(0)}{\partial\Psi_{\rm Q}^{(1)}\over\partial r}-{\dd g_{\rm Q}^{(0)}\over\dd r}\Psi_{\rm Q}^{(1)}\right]\,.
\end{equation}
Using the Helmholtz theorem, we can express
$f_{\rm Q}^{(1)}\,{\hat{\bf e}}_r=-\nabla\varphi_{\rm Q}^{(1)}-\nabla\times{\bf A}_{\rm Q}^{(1)}$,
with
\begin{eqnarray}\label{eq:fQphiA}
  \varphi_{\rm Q}^{(1)}&=&
  {\displaystyle
  \int_V{\nabla^\prime\cdot[f_{\rm Q}^{(1)}({\bf x}^\prime)\,{\hat{\bf e}}_{r^\prime}]
  \over 4\pi|{\bf x}-{\bf x}^\prime|}\,\dd^3 x^{\prime}
  -\,\oint_S {[f_{\rm Q}^{(1)}({\bf x}^\prime)\,{\hat{\bf e}}_{r^\prime}]\cdot{\hat{\bf n}^\prime}
  \over 4\pi|{\bf x}-{\bf x}^\prime|}\,\dd^2 x^{\prime}\,,} \\
  {\bf A}_{\rm Q}^{(1)}&=&
  {\displaystyle
  -\int_V{\nabla^\prime\times[f_{\rm Q}^{(1)}({\bf x}^\prime)\,{\hat{\bf e}}_{r^\prime}]
  \over 4\pi|{\bf x}-{\bf x}^\prime|}\,\dd^3 x^{\prime}
  -\,\oint_S {[f_{\rm Q}^{(1)}({\bf x}^\prime)\,{\hat{\bf e}}_{r^\prime}]\times{\hat{\bf n}^\prime}
  \over 4\pi|{\bf x}-{\bf x}^\prime|}\,\dd^2 x^{\prime}\,.}
\end{eqnarray}
If $f_{\rm Q}^{(1)}$ decay rapidly enough as $|{\bf x}^\prime|$ tends to infinity, then as the integral is extended to the
entire space, the surface terms will vanish.

Consequently, we have
\begin{equation}\label{eq:QuMONDIgQresult}
  {\bf g}_{\rm Q}=g_{\rm Q}^{(0)}\,{\hat{\bf e}}_r-\epsilon\,\nabla\left[\Psi^{(1)}_{\rm Q}+\varphi^{(1)}_{\rm Q}\right]+{\cal O}(\epsilon^2)\,,
\end{equation}
\begin{equation}\label{eq:QuMONDIgNresult}
  {\bf g}_{\rm N}=\mu^{(0)}g_{\rm Q}^{(0)}\,{\hat{\bf e}}_r-\epsilon\,\mu^{(0)}\left[\nabla\Psi_{\rm Q}^{(1)}
  +{\cal B}^{(0)}{\dd g_{\rm Q}^{(0)}\over\dd r}\Psi_{\rm Q}^{(1)}\,{\hat{\bf e}}_r\right]+{\cal O}(\epsilon^2)\,.
\end{equation}
The mass distribution is
\begin{equation}\label{eq:QuMONDIrho0}
  4\pi G\rho^{(0)}=-\,{1\over r^2}{\partial r^2 g_N^{(0)}\over\partial r}
  =-\,{1\over r^2}{\partial r^2 \mu^{(0)}g_{\rm Q}^{(0)}\over\partial r}\,,
\end{equation}
and
\begin{equation}\label{eq:QuMONDIrho1}
  4\pi G \rho^{(1)}=\nabla\cdot\left\{\mu^{(0)}\left[\nabla\Psi_{\rm Q}^{(1)}
  +{\cal B}^{(0)}{\dd g_{\rm Q}^{(0)}\over\dd r}\Psi_{\rm Q}^{(1)}\,{\hat{\bf e}}_r\right]\right\}\,.
\end{equation}

\subsubsection{QuMOND II: $\nabla\times{\bf g}_{\rm Q}=\nabla\times\left(\nu{\bf g}_{\rm N}+\nabla\times{\bf A}\right)=0$}
\label{sec:QuMONDII}

Once again, the mathematical procedure is the same as in Section~\ref{sec:AQUALII}.
All we need to do is to change ${\bf g}_{\rm A}$ to ${\bf g}_{\rm N}$, ${\bf h}$ to $-{\bf A}$, ${\tilde{\mu}}$ to $\nu$, etc.
We start from
\begin{equation}\label{eq:QuMONDIIgN}
  {\bf g}_{\rm N}=g_{\rm N}^{(0)}\,{\hat{\bf e}}_r+\epsilon\, {\bf g}_{\rm N}^{(1)}+{\cal O}(\epsilon^2)
  =g_{\rm N}^{(0)}\,{\hat{\bf e}}_r-\epsilon\, \nabla\Phi_{\rm N}^{(1)}+{\cal O}(\epsilon^2)\,,
\end{equation}
where $\Phi_{\rm N}^{(1)}(r,\theta,\phi)$ can be called the ``deformation potential''.
The compatibility condition (Equation~(\ref{eq:QuMONDcompatibleQ})) becomes
\begin{equation}\label{eq:QuMONDIIcompatible}
  0=\nu^{(0)}{\cal A}^{(0)}\left[g_{\rm N}^{(0)}\nabla g_{{\rm N}r}^{(1)}
  -{\dd g_{\rm N}^{(0)}\over\dd r}{\bf g}_{\rm N}^{(1)}\right]\times{\hat{\bf e}}_r
  +\nabla\times\nabla\times{\bf A}_{\rm N}^{(1)}\,,
  \quad
  {\cal A}^{(0)}=\left[{1\over a_0\nu}{\dd\nu\over\dd x_{\rm N}}\right]^{(0)}\,.
\end{equation}
Solving Equation~(\ref{eq:QuMONDIIcompatible}) implies
\begin{equation}\label{eq:QuMONDIIqN}
  q_{\rm N}^{(1)}\,{\hat{\bf e}}_r=\nabla\Upsilon_{\rm N}^{(1)}+\nabla\times{\bf A}_{\rm N}^{(1)}\,,
  \quad
  q_{\rm N}^{(1)}=\nu^{(0)}{\cal A}^{(0)}
  \left[g_{\rm N}^{(0)}{\partial\Phi^{(1)}_{\rm N}\over\partial r}-{\dd g_{\rm N}^{(0)}\over\dd r}\Phi^{(1)}_{\rm N}\right]\,,
\end{equation}
and by the Helmholtz theorem
\begin{eqnarray}\label{eq:qNUpsilonA}
  \Upsilon^{(1)}_{\rm N}&=&
  {\displaystyle
  -\int_V{\nabla^\prime\cdot[q_{\rm N}^{(1)}({\bf x}^\prime)\,{\hat{\bf e}}_{r^\prime}]
  \over 4\pi|{\bf x}-{\bf x}^\prime|}\,\dd^3 x^{\prime}
  +\,\oint_S {[q_{\rm N}^{(1)}({\bf x}^\prime)\,{\hat{\bf e}}_{r^\prime}]\cdot{\hat{\bf n}^\prime}
  \over 4\pi|{\bf x}-{\bf x}^\prime|}\,\dd^2 x^{\prime}\,,} \\
  {\bf A}^{(1)}_{\rm N}&=&
  {\displaystyle
  \int_V{\nabla^\prime\times[q_{\rm N}^{(1)}({\bf x}^\prime)\,{\hat{\bf e}}_{r^\prime}]
  \over 4\pi|{\bf x}-{\bf x}^\prime|}\,\dd^3 x^{\prime}
  +\,\oint_S {[q_{\rm N}^{(1)}({\bf x}^\prime)\,{\hat{\bf e}}_{r^\prime}]\times{\hat{\bf n}^\prime}
  \over 4\pi|{\bf x}-{\bf x}^\prime|}\,\dd^2 x^{\prime}\,.}
\end{eqnarray}
If $q_{\rm N}^{(1)}$ decay rapidly enough as $|{\bf x}^\prime|$ tends to infinity, then as the integral is extended to the
entire space, the surface terms will vanish.

Consequently, we have
\begin{eqnarray}\label{eq:QuMONDIIgQresult}
  {\bf g}_{\rm Q}&=&\nu^{(0)}g_{\rm N}^{(0)}\,{\hat{\bf e}}_r +\epsilon\, \left\{\nu^{(0)}
  \left[{\bf g}_{\rm N}^{(1)}+{\cal A}^{(0)}g_{\rm N}^{(0)}g_{{\rm N}r}^{(1)}\,{\hat{\bf e}}_r\right]
  +\nabla\times{\bf A}_{\rm N}^{(1)}\right\}+{\cal O}(\epsilon^2) \nonumber \\
  &=&\nu^{(0)}g_{\rm N}^{(0)}\,{\hat{\bf e}}_r-\epsilon\,\left\{\nu^{(0)}\left[\nabla\Phi^{(1)}_{\rm N}
  +{\cal A}^{(0)}{\dd g_{\rm N}^{(0)}\over\dd r}\Phi^{(1)}_{\rm N}\,{\hat{\bf e}}_r\right]+\nabla\Upsilon_{\rm N}^{(1)}\right\}
  +{\cal O}(\epsilon^2)\,.
\end{eqnarray}
The mass distribution is
\begin{equation}\label{eq:QuMONDIIrho0}
  4\pi G\rho^{(0)}=-\,{1\over r^2}{\partial r^2 g_N^{(0)}\over\partial r}\,,
\end{equation}
and
\begin{equation}\label{eq:QuMONDIIrho1}
  4\pi G \rho^{(1)}=\nabla^2\Phi_{\rm N}^{(1)}\,.
\end{equation}

\section{An example}\label{sec:example}

In this section, we present a simple example in AQUAL I.

First, it is interesting to point out the following:
\begin{itemize}
\item if $\Psi^{(1)}_{\rm N}(r,\theta,\phi)= g_{\rm N}^{(0)}(r)\psi^{(1)}_{\rm N}(\theta,\phi)$ in AQUAL I
or $\Phi^{(1)}_{\rm A}(r,\theta,\phi)= g_{\rm A}^{(0)}(r)\phi^{(1)}_{\rm A}(\theta,\phi)$ in AQUAL II,
then ${\tilde{\mu}}{\bf g}_{\rm A}={\bf g}_{\rm N}$ or ${\bf g}_{\rm A}={\tilde{\nu}}{\bf g}_{\rm N}$
up to first order (i.e., $\nabla\times{\bf h}={\cal O}(\epsilon^2)$);
\item if $\Psi_{\rm Q}^{(1)}(r,\theta,\phi)= g_{\rm Q}^{(0)}(r)\psi_{\rm Q}^{(1)}(\theta,\phi)$ in QuMOND I
or $\Phi^{(1)}_{\rm N}(r,\theta,\phi)= g_{\rm N}^{(0)}(r)\phi^{(1)}_{\rm N}(\theta,\phi)$ in QuMOND II,
then ${\mu}{\bf g}_{\rm Q}={\bf g}_{\rm N}$ or ${\bf g}_{\rm Q}={\nu}{\bf g}_{\rm N}$
up to first order (i.e., $\nabla\times{\bf A}={\cal O}(\epsilon^2)$).
\end{itemize}
That is, if one of these conditions is satisfied, then the problem becomes similar to spherical ones.
Moreover, the corresponding ``deformation potential'' of AQUAL I, $\Psi^{(1)}_{\rm N}$, does not depend on the interpolation function
if $\rho^{(1)}$ is given, see Equation~(\ref{eq:AQUALIrho1}).
(From Equation~(\ref{eq:QuMONDIIrho1}), we note that if $\rho^{(1)}$ is given, then the ``deformation potential'' of QuMOND II,
$\Phi^{(1)}_{\rm N}$, does not depend on the interpolation function in general.)

For simplicity, we take the Bekenstein form ($(\alpha,\eta)=(1,0)$ in Equation~(\ref{eq:canonicalnu})) in AQUAL I,
and $\Psi^{(1)}_{\rm N}(r,\theta,\phi)= g_{\rm N}^{(0)}(r)\psi^{(1)}_{\rm N}(\theta,\phi)$, then Equation~(\ref{eq:AQUALIgAresult}) gives
\begin{equation}\label{eq:specificgr}
  g_{{\rm A}r}=\left(1+x^{-1/2}_{\rm N}\right)^{(0)}g_{\rm N}^{(0)}
  -\epsilon\,\left(1+\half x^{-1/2}_{\rm N}\right)^{(0)}{\dd g_{\rm N}^{(0)}\over\dd r}\psi^{(1)}_{\rm N}\,,
\end{equation}
\begin{equation}\label{eq:specificgtheta}
  g_{{\rm A}\theta}=-\epsilon\,\left(1+x^{-1/2}_{\rm N}\right)^{(0)}\left({g_{\rm N}^{(0)}\over r}\right)
  {\partial\psi^{(1)}_{\rm N}\over\partial\theta_{\rm l}}\,,
\end{equation}
\begin{equation}\label{eq:specificgphi}
  g_{{\rm A}\phi}=-\epsilon\,\left(1+x^{-1/2}_{\rm N}\right)^{(0)}\left({g_{\rm N}^{(0)}\over r\sin\theta_{\rm l}}\right)
  {\partial\psi^{(1)}_{\rm N}\over\partial\phi_{\rm l}}\,.
\end{equation}
The density is given by
$4\pi G\rho=\nabla^2\Phi_{\rm N}
=-\nabla\cdot\left(g_{\rm N}^{(0)}\,{\hat{\bf e}}_r\right)+\epsilon\,\nabla^2\left(g_{\rm N}^{(0)}\psi^{(1)}_{\rm N}\right)$.

To educate ourselves, here is a simple example that gives a flattened axial symmetric mass distribution
(oblate-like distribution: $\sin^2\theta$),
\begin{equation}\label{eq:examplegNpsiN}
  g_{\rm N}^{(0)}=-\,{Gm_0\over r_0^2}\left(r\over r_0\right)^p\,,
  \quad
  \psi_{\rm N}^{(1)}=r_0\left[p(p+1)\cos^2\theta-(p^2+p-4)\right]\,,
\end{equation}
where $-2<p<0$.
Substituting Equation~(\ref{eq:examplegNpsiN}) into Equations~(\ref{eq:specificgr})--(\ref{eq:specificgphi}) explicitly provides ${\bf g}_{\rm A}$,
and the corresponding potential (${\bf g}_{\rm A}=-\nabla\Phi_{\rm A}$)
\begin{eqnarray}\label{eq:specificpotential}
  \Phi_{\rm A}&=&{Gm_0\over r_0}\left\{\left[{1\over\left(p+1\right)}\left({r\over r_0}\right)^{(p+1)}
  +{2\over\left(p+2\right)}\sqrt{{a_0r_0^2\over Gm_0}\,}\left({r\over r_0}\right)^{(p+2)/2}\right]\right. \nonumber \\
  &&\left.-\,\epsilon\,\left[\left(r\over r_0\right)^p+\sqrt{{a_0r_0^2\over Gm_0}\,}\left(r\over r_0\right)^{p/2}\right]
  \left[p(p+1)\cos^2\theta-(p^2+p-4)\right]\right\}\,.
\end{eqnarray}
Putting $a_0=0$ in Equation~(\ref{eq:specificpotential}) gives $\Phi_{\rm N}$.
Moreover, the density is
\begin{equation}\label{eq:examplerho}
  \rho={m_0\over 4\pi r_0^3}\left[(p+2)\left(r\over r_0\right)^{p-1}
  +\epsilon\,p(p-2)(p+1)(p+3)\left(r\over r_0\right)^{p-2}\sin^2\theta\right]\,.
\end{equation}
To ensure a positive density, we should take $\epsilon<0$ for $-2<p<-1$, and $\epsilon>0$ for $-1<p<0$.
Nevertheless, the model has the shortcoming that the density is dominated by the first-order term at small $r$.

Suppose the axis of symmetry is perpendicular to the line of sight.
Set up a Cartesian coordinate system in the observer frame $(\xi,\eta,\zeta)$ such that the line of sight is along the $\zeta$-axis,
and the axis of symmetry of the object is along the $\xi$-axis.
Thus, we have $r^2=\xi^2+\eta^2+\zeta^2$ and $\cos\theta=\xi/r$, and
\begin{eqnarray}\label{eq:specificgAxi}
  g_{{\rm A}\xi}&=&-\,{Gm_0\over r_0^2}\left({\xi\over r}\right)\left\{\left[\left({r\over r_0}\right)^p
  +\sqrt{{a_0r_0^2\over Gm_0}\,}\left({r\over r_0}\right)^{p/2}\right]\right. \nonumber \\
  &&\left.+\,\epsilon\,p\left[(p-3)(p+2)\left(r\over r_0\right)^{p-1}
  +{\left(p^2-3p-8\right)\over 2}\sqrt{{a_0r_0^2\over Gm_0}\,}\left(r\over r_0\right)^{(p-2)/2}\right] \right. \nonumber \\
  &&\left.-\,\epsilon\,p(p+1)\left({\xi^2\over r^2}\right)\left[(p-2)\left(r\over r_0\right)^{p-1}
  +{\left(p-4\right)\over 2}\sqrt{{a_0r_0^2\over Gm_0}\,}\left(r\over r_0\right)^{(p-2)/2}\right]\right\}\,,
\end{eqnarray}
\begin{eqnarray}\label{eq:specificgAeta}
  g_{{\rm A}\eta}&=&-\,{Gm_0\over r_0^2}\left({\eta\over r}\right)\left\{\left[\left({r\over r_0}\right)^p
  +\sqrt{{a_0r_0^2\over Gm_0}\,}\left({r\over r_0}\right)^{p/2}\right]\right. \nonumber \\
  &&\left.+\,\epsilon\,p\left(p^2+p-4\right)\left[\left(r\over r_0\right)^{p-1}
  +{1\over 2}\sqrt{{a_0r_0^2\over Gm_0}\,}\left(r\over r_0\right)^{(p-2)/2}\right] \right. \nonumber \\
  &&\left.-\,\epsilon\,p(p+1)\left({\xi^2\over r^2}\right)\left[(p-2)\left(r\over r_0\right)^{p-1}
  +{\left(p-4\right)\over 2}\sqrt{{a_0r_0^2\over Gm_0}\,}\left(r\over r_0\right)^{(p-2)/2}\right]\right\}\,.
\end{eqnarray}

For illustration purposes, we apply this model to strong gravitational lensing (see Appendix~\ref{sec:lens}).
Figure~\ref{fig:figure1} shows the critical curves and caustics for the case $p=-3/2$.

\section{Summary and discussion}\label{sec:discussion}

Non-spherical systems in the framework of MOND are a lot more difficult to analyse than
spherical systems.
There are plenty of astrophysical objects that can be approximated by a slightly deformed spherical distribution.
As an alternative to dark matter, it is desirable to develop methods or algorithms to deal with such systems.
Based on the compatibility condition, we propose a method to analyse slightly deformed spherical systems (i.e., slightly non-spherical systems)
in the framework of MOND.
There are two formulations of MOND, namely, AQUAL and QuMOND, and the compatibility condition can be written in two ways,
and hence we have four different approaches, see Sections~\ref{sec:AQUALI}--\ref{sec:QuMONDII}.
In general, this involves solving the corresponding ``deformation potential'' when the mass distribution is given.

To examine the dynamics of an object, in principle, one requires observations of the distribution of its mass (e.g., brightness distribution)
and the gravitational acceleration of the object (e.g., velocity distribution or light bending in gravitational lensing).
For data fitting, one may start from a mass model and compute the gravitational acceleration,
or the other way round.
Here, we briefly summarise these two approaches in our proposed method.


\begin{itemize}
\item[(1)] {\it Start from a model of mass distribution.}
\item[] Suppose we have a model of mass distribution $\rho=\rho^{(0)}(r)+\epsilon\,\rho^{(1)}(r,\theta,\phi)$,
and a prescribed interpolation function, then we can deduce the gravitational acceleration as follows.
\item{AQUAL I} with prescribed $\tilde{\nu}$
  \begin{itemize}
  \item Equation~(\ref{eq:AQUALIrho0}) can be integrated to give the zeroth-order Newtonian acceleration $g_{\rm N}^{(0)}$;
  \item Equation~(\ref{eq:AQUALIrho1}) becomes a differential equation for the ``deformation potential'' $\Psi_{\rm N}^{(1)}$;
  \item once $\Psi_{\rm N}^{(1)}$ is known, the MONDian acceleration ${\bf g}_{\rm A}$ is given by Equation~(\ref{eq:AQUALIgAresult}).
  \end{itemize}
\item{AQUAL II} with prescribed $\tilde{\mu}$
  \begin{itemize}
  \item Equation~(\ref{eq:AQUALIIrho0}) can be integrated to give the zeroth-order MONDian acceleration $g_{\rm A}^{(0)}$;
  \item Equation~(\ref{eq:AQUALIIrho1}) becomes a differential equation for the ``deformation potential'' $\Phi_{\rm A}^{(1)}$;
  \item once $\Phi_{\rm A}^{(1)}$ is known, the MONDian acceleration ${\bf g}_{\rm A}$ is given by Equation~(\ref{eq:AQUALIIgA}).
  \end{itemize}
\item{QuMOND I} with prescribed $\mu$
  \begin{itemize}
  \item Equation~(\ref{eq:QuMONDIrho0}) can be integrated to give the zeroth-order MONDian acceleration $g_{\rm Q}^{(0)}$;
  \item Equation~(\ref{eq:QuMONDIrho1}) becomes a differential equation for the ``deformation potential'' $\Psi_{\rm Q}^{(1)}$;
  \item once $\Psi_{\rm Q}^{(1)}$ is known, we can solve $\nabla^2\varphi_{\rm Q}^{(1)}=-\nabla\cdot\left[f_{\rm Q}^{(1)}{\hat{\bf e}}_r\right]$
  for $\varphi_{\rm Q}^{(1)}$, where $f_{\rm Q}^{(1)}$ is given by Equation~(\ref{eq:QuMONDIFQ});
  \item once $\Psi_{\rm Q}^{(1)}$ and $\varphi_{\rm Q}^{(1)}$ are known,
  the MONDian acceleration ${\bf g}_{\rm Q}$ is given by Equation~(\ref{eq:QuMONDIgQresult}).
  \end{itemize}
\item{QuMOND II} with prescribed $\nu$
  \begin{itemize}
  \item Equation~(\ref{eq:QuMONDIIrho0}) can be integrated to give the zeroth-order Newtonian acceleration $g_{\rm N}^{(0)}$;
  \item Equation~(\ref{eq:QuMONDIIrho1}) becomes a differential equation for the ``deformation potential'' $\Phi_{\rm N}^{(1)}$;
  \item once $\Phi_{\rm N}^{(1)}$ is known, we can solve $\nabla^2\Upsilon_{\rm N}^{(1)}=\nabla\cdot\left[q_{\rm N}^{(1)}\,{\hat{\bf e}}_r\right]$
  for $\Upsilon_{\rm N}^{(1)}$, where $q_{\rm N}^{(1)}$ is given by Equation~(\ref{eq:QuMONDIIqN});
  \item once $\Phi_{\rm N}^{(1)}$ and $\Upsilon_{\rm N}^{(1)}$ are known,
  the MONDian acceleration ${\bf g}_{\rm A}$ is given by Equation~(\ref{eq:QuMONDIIgQresult}).
  \end{itemize}
\end{itemize}

\begin{itemize}
\item[(2)] {\it Start from a model of gravitational acceleration.}
\item[] If we start from a model of acceleration ${\bf g}=g^{(0)}(r)\,{\hat{\bf e}}_r+\epsilon\,{\bf g}^{(1)}(r,\theta,\phi)$,
and a prescribed interpolation function, then we can deduce the mass distribution. However, in general, there is no guarantee that
the deduced density is non-negative everywhere.
\item{AQUAL} (either I or II) with prescribed $\tilde{\mu}$
  \begin{itemize}
  \item the density is given by $4\pi G\rho=-\nabla\cdot\left({\tilde{\mu}}{\bf g}\right)$.
  \end{itemize}
\item{QuMOND I} with prescribed $\mu$
  \begin{itemize}
  \item set $g_{\rm Q}^{(0)}=g^{(0)}$, then the zeroth-order density $\rho^{(0)}$ is given by Equation~(\ref{eq:QuMONDIrho0});
  \item Equation~(\ref{eq:QuMONDIGQ}) gives $\nabla\cdot{\bf F}_{\rm Q}^{(1)}=\nabla\cdot{\bf g}^{(1)}$;
  \item this equation together with Equation~(\ref{eq:QuMONDIFQ}) give a differential equation for the ``deformation potential'' $\Psi_{\rm Q}^{(1)}$;
  \item once $\Psi_{\rm Q}^{(0)}$ is known, the first-order density $\rho^{(1)}$ is given by Equation~(\ref{eq:QuMONDIrho1}).
  \end{itemize}
\item{QuMOND II} with prescribed $\nu$
  \begin{itemize}
  \item solve $\nu^{(0)} g_{\rm N}^{(0)}=g^{(0)}$ for $g_{\rm N}^{(0)}$ (or $g_{\rm N}^{(0)}=\mu^{(0)} g^{(0)}$ if $\mu$ is given);
  \item the zeroth-order density $\rho^{(0)}$ is the given by Equation~(\ref{eq:QuMONDIIrho0});
  \item $\nabla\cdot{\bf g}_{\rm Q}^{(1)}=\nabla\cdot{\bf g}^{(1)}$ together with Equation~(\ref{eq:QuMONDIIgQresult})
  give a differential equation for the ``deformation potential'' $\Phi_{\rm N}^{(1)}$;
  \item once $\Phi_{\rm N}^{(1)}$ is known, the first-order density $\rho^{(1)}$ is given by Equation~(\ref{eq:QuMONDIIrho1}).
  \end{itemize}
\end{itemize}

In most cases, we will follow the first procedure as it is rather straightforward to model the mass distribution from
the surface brightness distribution.

For some specific forms of the deformation potential,
e.g., $\Psi^{(1)}_{\rm N}(r,\theta,\phi)= g_{\rm N}^{(0)}(r)\psi^{(1)}_{\rm N}(\theta,\phi)$ in AQUAL I,
$\Phi^{(1)}_{\rm A}(r,\theta,\phi)= g_{\rm A}^{(0)}(r)\phi^{(1)}_{\rm A}(\theta,\phi)$ in AQUAL II,
$\Psi_{\rm Q}^{(1)}(r,\theta,\phi)= g_{\rm Q}^{(0)}(r)\psi_{\rm Q}^{(1)}(\theta,\phi)$ in QuMOND I, or
$\Phi^{(1)}_{\rm N}(r,\theta,\phi)= g_{\rm N}^{(0)}(r)\phi^{(1)}_{\rm N}(\theta,\phi)$ in QuMOND II,
the Newtonian and MONDian gravitational accelerations are in the same direction (up to first order in $\epsilon$),
i.e., ${\bf g}_{\rm A}\parallel{\bf g}_{\rm N}$, or ${\bf g}_{\rm Q}\parallel{\bf g}_{\rm N}$.
Spherical systems have the same property.

We would like to point out an attractive feature of QuMOND II.
For a prescribed $\rho^{(1)}$ the ``deformation potential'' $\Phi^{(1)}_{\rm N}$ of QuMOND II
does not depend on the interpolation function.
A similar feature occurs in AQUAL I, but only if the condition
$\Psi^{(1)}_{\rm N}(r,\theta,\phi)= g_{\rm N}^{(0)}(r)\psi^{(1)}_{\rm N}(\theta,\phi)$ is satisfied
(see Equation~(\ref{eq:AQUALIrho1})).

As an alternative to the dark matter paradigm, many of MOND's studies were devoted to galaxy systems.
The luminous parts of these systems (either elliptical galaxies, spiral galaxies, or clusters of galaxies), in general, are asymmetric.
The procedure described above provides a tool enabling us to analyse many aspherical systems in the framework of MOND
(it cannot address every asymmetric configuration though).
The method is flexible enough for us to perform some serious modelling on the (baryonic) mass distribution of galaxy systems
(in particular, elliptical galaxies) when we study phenomena such as gravitational lensing and stellar dynamics in these systems.

Cold dark matter simulations showed that the shape of the dark matter haloes is in generally aspherical
\citep[their orientations with respect to the baryonic matter in galaxies are studied as well; see, e.g.,][]
{Jing2002,Springel2005,Hayashi2007,VeraCiro2011,VeraCiro2014,Schneider2012,Velliscig2015,Gerhard2013}.
Observations such as gravitational lensing, or stellar and satellites kinematics, may place constraints on the shape of the haloes
and the corresponding gravitational field
\citep[e.g.,][]{Bailin2008,Howell2010,Deason2011,Bett2012,vanUitert2012,Hayashi2012,Hayashi2014,Hayashi2015,
VeraCiro2013,Joachimi2013a,Joachimi2013b,Schrabback2015}.
These observations also place constraints on MOND as well.
The MONDian gravitational field is dictated by $\rho_{\rm b}$,
the shape and mass distribution of the baryons in the galaxy (the luminous part of the galaxy).
Suppose a Newtonian field equivalent to the MONDian field is produced by an effective total mass $\rho^\prime_{\rm t}$,
then $\rho^\prime_{\rm DM}=\rho^\prime_{\rm t}-\rho_{\rm b}$ can be called the effective dark matter distribution.
In QuMOND formulation, the expression for the effective dark matter distribution is simple,
as it involves the Newtonian field from the baryons only:
\begin{equation}\label{eq:QuMONDDM}
  4\pi G\rho^\prime_{\rm DM,Q}=-\,{\dd\nu\over\dd|{\bf g}_{\rm N}|}\,{\bf g}_{\rm N}\cdot\nabla\left|{\bf g}_{\rm N}\right|\,,
\end{equation}
where $\nabla\cdot{\bf g}_{\rm N}=-4\pi G\rho_{\rm b}$, and $\nu$ depends on $|{\bf g}_{\rm N}|$.
In the AQUAL formulation
\begin{equation}\label{eq:AQUALDM}
  4\pi G\rho^\prime_{\rm DM,A}={\dd{\tilde\mu}\over\dd|{\bf g}_{\rm A}|}\,{\bf g}_{\rm A}\cdot\nabla\left|{\bf g}_{\rm A}\right|\,,
\end{equation}
where ${\bf g}_{\rm A}$ and ${\bf g}_{\rm N}$ are related by Equation~(\ref{eq:AQUALacc}) or (\ref{eq:AQUALinvert}), and
${\tilde\mu}$ depends on $|{\bf g}_{\rm A}|$.
Observations will place a constraint on the interpolation function ${\tilde\mu}(|{\bf g}_{\rm A}|/a_0)$ or $\nu(|{\bf g}_{\rm N}|/a_0)$,
and may even distinguish between AQUAL and QuMOND.

As illustrated in Section~\ref{sec:example} (and Appendix~\ref{sec:lens}),
strong gravitational lensing will be a straightforward application of the method presented in this article.
A deformed spherical lens can be used to study arcs, rings, quadruple-image systems in gravitational lens surveys, such as
CASTLES,
SLACS,
Master Lens,
SQLS,
CLASS,
CLASH,
GLASS,
etc..
In particular, quadruple-image systems with high-quality data are nice targets, such as
B1422+231 \citep[][]{Nierenberg2014},
B1608+656 \citep[][]{Suyu2009},
HE0435-1223 \citep[][]{Kochanek2006},
MGJ0414+0534 \citep[][]{Trotter2000}, and
PG1115+080 \citep[][]{Impey1998}.
The results will place some constraints on the parameters in MOND and/or the Hubble constant.
We will consider analysis of these systems elsewhere.

\acknowledgments
The author is grateful to Yong Tian and Mu-Chen Chiu for stimulating discussions on the development of this work.
This work is supported in part by the Taiwan Ministry of Science and Technology,
grants MOST 102-2112-M-008-019-MY3 and MOST 104-2923-M-008-001-MY3.

\appendix

\section{Gravitational redshift}\label{sec:redshift}

In this appendix, we use simple estimates to place some constraint on the MOND interpolation function $\tilde\mu(x)$
or its inverse $\tilde\nu(x_{\rm N})$ by the gravitational redshift measurement from a tabletop atomic interferometer experiment \citep[][]{Muller2010}.

Assuming that the Earth is a sphere, we have
for AQUAL $\tilde{\mu}(x_{\rm A}){\bf\nabla}\Phi_{\rm A}={\bf\nabla}\Phi_{\rm N}$ and
${\bf\nabla}\Phi_{\rm A}=\tilde{\nu}(x_{\rm N}){\bf\nabla}\Phi_{\rm N}$ (see Equation~(\ref{eq:AQUALsym})),
and for QuMOND $\nu(x_{\rm N}){\bf\nabla}\Phi_{\rm N}={\bf\nabla}\Phi_{\rm Q}$ and
${\bf\nabla}\Phi_{\rm N}=\mu(x_{\rm Q}){\bf\nabla}\Phi_{\rm Q}$
(see Equation~(\ref{eq:QuMONDsym})).
Here, $x_{\rm N}=|{\bf\nabla}\Phi_{\rm N}|/a_0$, $x_{\rm A}=|{\bf\nabla}\Phi_{\rm A}|/a_0$ and $x_{\rm Q}=|{\bf\nabla}\Phi_{\rm Q}|/a_0$.
The accumulated phase due to redshift can be expressed as \citep[see][]{Muller2010}
\begin{equation}\label{eq:phasedifference}
  \Delta\phi\approx -\,{\omega_{\rm C}\over c^2}\int z{\hat e}_r\cdot{\bf\nabla}\Phi_{\rm A}(r_{\rm E})\,\dd t
  =-\,{\omega_{\rm C}\over c^2}\int {\tilde\nu}z{\hat e}_r\cdot{\bf\nabla}\Phi_{\rm N}(r_{\rm E})\,\dd t\,,
\end{equation}
for AQUAL. Here, $\omega_{\rm C}=mc^2/\hbar$ is the Compton wavelength of the atom, $r_{\rm E}$ is the radius of the Earth,
and $r=r_{\rm E}+z$ ($z\ll r_{\rm E}$).
If we replace $\Phi_{\rm A}$ and ${\tilde\nu}$ by $\Phi_{\rm Q}$ and $\nu$ in Equation~(\ref{eq:phasedifference}),
then we obtain the corresponding expression for QuMOND.

The parameter used to model the anomalies in gravitational redshift in \citet{Muller2010} is
\begin{equation}\label{eq:redshiftparameter}
  \beta=\tilde{\nu}(x_{\rm N})-1=\left[1+\half\left(\sqrt{4\chi_{\rm N}^{-\alpha}+\eta^2}-\eta\right)\right]^{1/\alpha}-1\,,
\end{equation}
where we adopted the inverted interpolation function (Equation~(\ref{eq:canonicalnu}); note that $\alpha>0$ and $\eta \ge 0$).
Supposing that this canonical form is valid in the high acceleration regime (i.e., the Newtonian limit, $x_{\rm N}\gg 1$, ${\tilde\nu}-1\ll 1$),
then the experiment by \citet{Muller2010} would place some constraint on the canonical form.
Often, an upper bound of $\beta$ is obtained in gravitational redshift experiments.
Equation~(\ref{eq:redshiftparameter}) thus gives a constraint on $(\alpha,\eta)$,
\begin{equation}\label{eq:separation}
  \eta>{x_{\rm N}^{-\alpha}-\left[\left(1+\beta_{\rm u}\right)^\alpha-1\right]^2\over\left[\left(1+\beta_{\rm u}\right)^\alpha-1\right]}\,,
\end{equation}
where $\beta_{\rm u}$ is the observed upper bound of $\beta$.

In the atomic interferometer experiment, \citet{Muller2010} obtained $\beta_{\rm u}=7\times 10^{-9}$.
If we take the nominal value of gravitational acceleration on Earth's surface $|{\bf\nabla}\Phi_{\rm N}(r_{\rm E})|=9.81$ m s$^{-2}$
and the acceleration constant $a_0=1.2\times 10^{-10}$ m s$^{-2}$ \citep[e.g.,][]{Sanders2002}, then we have $x_{\rm N}=8.175\times 10^{10}$.
Figure~\ref{fig:figure2} shows the constraint on the parameter space $(\alpha,\eta)$.
The white (gray) region in the figure is the parameter space that is consistent with (excluded by) the experiment.
For instance, for $\eta=0$, $\alpha$ must be larger than 1.464; and for $\alpha=1$, $\eta$ must be larger than 0.001747.

\section{Lens equation}\label{sec:lens}

In this appendix, we write down the lens equation for the example in Section~\ref{sec:example}.
Assuming a small angle of deflection, the lens equation is
\begin{equation}\label{eq:lenseq}
  {\vec\beta}={\vec\vartheta}+{2 D_{\rm LS}\over c^2 D_{\rm S}}\int {\bf g}_\perp\,\dd\zeta\,,
\end{equation}
where ${\vec\beta}$ and ${\vec\vartheta}$ are the position angle of the source and image, respectively.
$D_{\rm LS}$ and $D_{\rm S}$ are the distances of the source from the lens and from the observer, respectively.
Here, the integration is taken along the undeflected path $\zeta$ from the source to the observer
(which can be consider as from negative infinity to positive infinity).
${\bf g}_\perp$ is the gravitational acceleration perpendicular to this path.
${\bf g}_\perp$ is a function of $(\xi,\eta,\zeta)$, and $\xi\approx D_{\rm L}\vartheta_\xi$, $\eta\approx D_{\rm L}\vartheta_\eta$,
where $D_{\rm L}$ is the distance between the lens and the observer.
Substituting Equations~(\ref{eq:specificgAxi}) and (\ref{eq:specificgAeta}) into Equation~(\ref{eq:lenseq}), we get
(with $\vartheta^2=\vartheta_\xi^2+\vartheta_\eta^2$)
\begin{eqnarray}\label{eq:lenseqspecificxi}
  \beta_\xi&=&\vartheta_\xi-{\vartheta_{\rm E}^2\vartheta_\xi\over\vartheta_{\rm m}^2}
  \left\{{\vartheta^p\over\vartheta_{\rm m}^p}\left[{\cal J}_1
  +{\epsilon\,\vartheta_{\rm m}\over\vartheta}\,\left({\cal J}_5-{\vartheta_\xi^2\over\vartheta^2}\,{\cal J}_3\right)\right]\right. \nonumber \\
  & &\quad\quad\left. +\,{\vartheta_{\rm m}\over\vartheta_0}\,{\vartheta^{p/2}\over\vartheta_{\rm m}^{p/2}}\left[{\cal J}_2
  +{\epsilon\,\vartheta_{\rm m}\over\vartheta}\,\left({\cal J}_6
  -{\vartheta_\xi^2\over\vartheta^2}\,{\cal J}_4\right)\right]\right\}\,,
\end{eqnarray}
\begin{eqnarray}\label{eq:lenseqspecificeta}
  \beta_\eta&=&\vartheta_\eta-{\vartheta_{\rm E}^2\vartheta_\eta\over\vartheta_{\rm m}^2}
  \left\{{\vartheta^p\over\vartheta_{\rm m}^p}\left[{\cal J}_1
  +{\epsilon\,\vartheta_{\rm m}\over\vartheta}\,\left({\cal J}_7-{\vartheta_\xi^2\over\vartheta^2}\,{\cal J}_3\right)\right]\right. \nonumber \\
  & &\quad\quad\left. +\,{\vartheta_{\rm m}\over\vartheta_0}\,{\vartheta^{p/2}\over\vartheta_{\rm m}^{p/2}}\left[{\cal J}_2+
  {\epsilon\,\vartheta_{\rm m}\over\vartheta}\,\left({\cal J}_8
  -{\vartheta_\xi^2\over\vartheta^2}\,{\cal J}_4\right)\right]\right\}\,,
\end{eqnarray}
where
\begin{equation}\label{eq:thEandth0}
  \vartheta_{\rm E}^2={4Gm_0D_{\rm LS}\over c^2D_{\rm S}D_{\rm L}}\,,
  \quad
  \vartheta_0^2={Gm_0\over D_{\rm L}^2 a_0}\,,
  \quad
  \vartheta_{\rm m}={r_0\over D_{\rm L}}\,,
\end{equation}
\begin{eqnarray}\label{eq:J1-8}
  {\cal J}_1&=&{\cal I}_{(1-p)/2}\,, \\
  {\cal J}_2&=&{\cal I}_{(2-p)/4}\,, \\
  {\cal J}_3&=&p(p-2)(p+1)\,{\cal I}_{(4-p)/2}\,, \\
  {\cal J}_4&=&\half p(p-4)(p+1)\,{\cal I}_{(8-p)/4}\,, \\
  {\cal J}_5&=&p(p-3)(p+2)\,{\cal I}_{(2-p)/2}\,, \\
  {\cal J}_6&=&\half p(p^2-3p-8)\,{\cal I}_{(4-p)/4}\,, \\
  {\cal J}_7&=&p(p^2+p-4)\,{\cal I}_{(2-p)/2}\,, \\
  {\cal J}_8&=&\half p(p^2+p-4)\,{\cal I}_{(4-p)/4}\,.
\end{eqnarray}
Here,
\begin{eqnarray}\label{eq:I}
  {\cal I}_q&=& {1\over 2}\int_{-{\tilde{D}_{\rm LS}}}^{\tilde{D}_{\rm L}^\prime} {\dd {\tilde\zeta}\over\left(1+{\tilde\zeta}^2\right)^q}
  ={1\over 2}\int_0^{\tilde{D}_{\rm L}^\prime} {\dd {\tilde\zeta}\over\left(1+{\tilde\zeta}^2\right)^q}
  +{1\over 2}\int_0^{\tilde{D}_{\rm LS}} {\dd {\tilde\zeta}\over\left(1+{\tilde\zeta}^2\right)^q} \nonumber \\
  &=&\half\left[{\tilde\zeta}\, _2F_1\left(\half,q,\threehalf,-{\tilde\zeta}^2\right)\right]_0^{\tilde{D}_{\rm L}^\prime}
  +\half\left[{\tilde\zeta}\, _2F_1\left(\half,q,\threehalf,-{\tilde\zeta}^2\right)\right]_0^{\tilde{D}_{\rm LS}} \nonumber \\
  &\approx& {\Gamma(\half-q)\over 4\Gamma(\threehalf-q)}\left({\tilde{D}_{\rm L}}^{\prime\,^{1-2q}}+{\tilde{D}_{\rm LS}}^{1-2q}\right)
  +{\sqrt{\pi}\,\Gamma(q-\half)\over 2\Gamma(q)}+\cdots\,,
\end{eqnarray}
where $\tilde\zeta=\zeta/\sqrt{\xi^2+\eta^2}$, $\tilde{D}_{\rm L}^\prime=D_{\rm L}^\prime/\sqrt{\xi^2+\eta^2}$
and $\tilde{D}_{\rm LS}=D_{\rm LS}/\sqrt{\xi^2+\eta^2}$, and note that $\tilde{D}_{\rm L}^\prime\,, \tilde{D}_{\rm LS}\gg 1$.
For $q>\half$, the first term in Equation~(\ref{eq:I}) is subordinate to the second term and can be neglected.
As we are only interested in $-2<p<0$, all ${\cal J}$s satisfy $q>\half$.

Equations~(\ref{eq:lenseqspecificxi}) and (\ref{eq:lenseqspecificeta})
are the mapping of the image to the source.
The determinant of the inverse of the Jacobian of the mapping gives the magnification of the image.
The positions of the image when the magnification becomes infinite form the so called critical lines.
The corresponding source positions form caustics.

For completeness, we write down the time-delay function:
\begin{equation}\label{eq:timedelay}
  \Delta T={(1+z_{\rm L})\over c}{D_{\rm L}D_{\rm S}\over D_{\rm LS}}
  \left[\half\left({\vec\theta}-{\vec\beta}\right)^2-\psi\right]\,,
\end{equation}
where $\psi=2D_{\rm LS}/(c^2D_{\rm L}D_{\rm S})\,\int_{-D_{\rm LS}}^{D_{\rm L}^\prime}\,\Phi\,\dd\zeta$.
Taking $\Phi_{\rm A}$ in Equation~(\ref{eq:specificpotential}) as $\Phi$, we have
\begin{eqnarray}\label{eq:timedelaypotential}
  \psi&=&\vartheta_{\rm E}^2
  \left\{{\vartheta^{p+2}\over\vartheta_{\rm m}^{p+2}}\left[{\cal J}_9
  +{\epsilon\,\vartheta_{\rm m}\over\vartheta}\,\left({\cal J}_{13}
  -{\vartheta_\xi^2\over\vartheta^2}\,{\cal J}_{11}\right)\right]\right. \nonumber \\
  & &\quad\quad \left. +\,{\vartheta_{\rm m}\over\vartheta_0}\,{\vartheta^{(p+4)/2}\over\vartheta_{\rm m}^{(p+4)/2}}\left[{\cal J}_{10}
  +{\epsilon\,\vartheta_{\rm m}\over\vartheta}\,\left({\cal J}_{14}
  -{\vartheta_\xi^2\over\vartheta^2}\,{\cal J}_{12}\right)\right]\right\}\,,
\end{eqnarray}
where
\begin{eqnarray}\label{eq:J9-14}
  {\cal J}_9&=&{1\over(p+1)}\,{\cal I}_{-(1+p)/2}\,, \\
  {\cal J}_{10}&=&{2\over(p+2)}\,{\cal I}_{-(2+p)/4}\,, \\
  {\cal J}_{11}&=&p(p+1)\,{\cal I}_{(2-p)/2}\,, \\
  {\cal J}_{12}&=&p(p+1)\,{\cal I}_{(4-p)/4}\,, \\
  {\cal J}_{13}&=&(p^2+p-4)\,{\cal I}_{-p/2}\,, \\
  {\cal J}_{14}&=&(p^2+p-4)\,{\cal I}_{-p/4}\,.
\end{eqnarray}

Moreover, the projected surface density $\Sigma=\int_{-\infty}^\infty \rho\,\dd\zeta$ is
\begin{equation}\label{eq:surfacedensity}
  \Sigma={m_0\over 2\pi r_0^2}\,{\vartheta^p\over\vartheta_{\rm m}^p}\left\{(p+2)\,{\cal I}_{(1-p)/2}
  +\,{\epsilon\,\vartheta_{\rm m}\over\vartheta}\,p(p-2)(p+1)(p+3)\,\left[{\cal I}_{(2-p)/2}
  -{\vartheta_\xi^2\over\vartheta^2}\,{\cal I}_{(4-p)/2}\right]\right\}\,.
\end{equation}

\begin{figure}
\epsscale{1.0}
\plotone{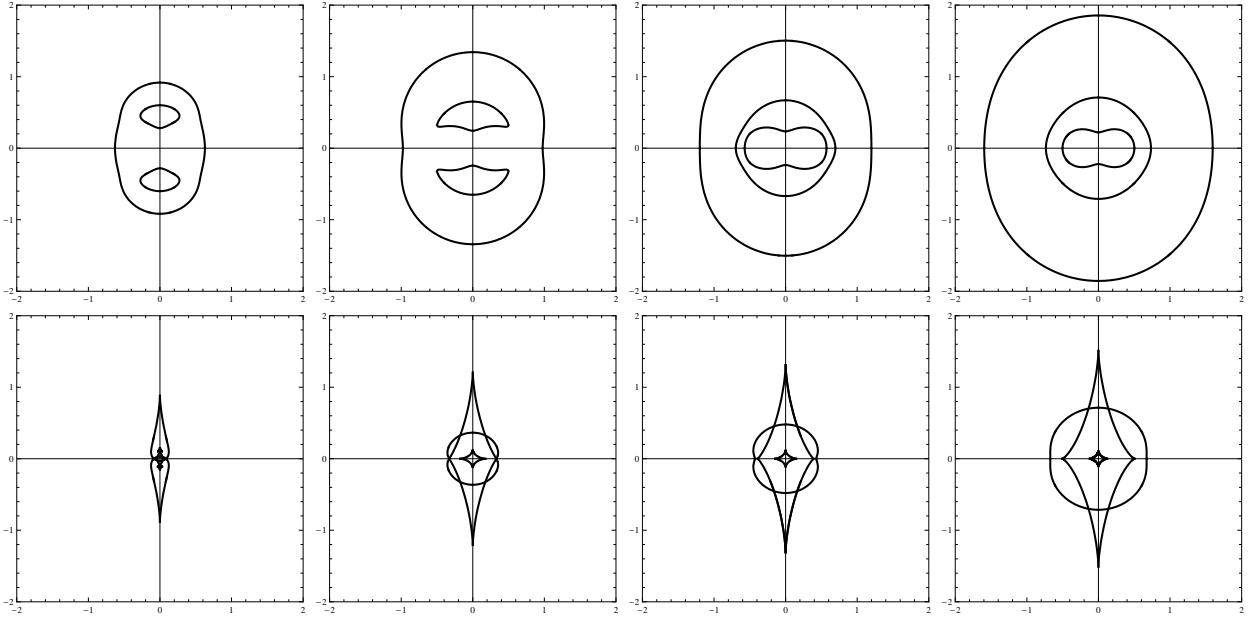}
\caption{
Example of strong gravitational lensing in AQUAL I.
Critical curves and caustics for the case $p=-3/2$ (i.e., $g_{\rm N}^{(0)}\propto r^{-3/2}$).
The upper row is the critical curves and the lower row is the caustics.
From left to right the parameters
$(\epsilon,\vartheta_{\rm E},\vartheta_0)$
for the columns are
$(-0.1,1,\infty)$, $(-0.1,1,4)$, $(-0.1,1,3)$, and $(-0.1,1,2)$.
The definitions of the parameters are given in Appendix~\ref{sec:lens}.
$\vartheta_0$ represents the ratio of the characteristic acceleration of the system to $a_0$.
The larger is $\vartheta_0$ the closer is the system to the Newtonian regime (cf. $x_{\rm A}$ in Equation~(1)).
The plots are in units of $\vartheta_{\rm m}=r_0/D_{\rm L}$ ($D_{\rm L}$ is the distance between the lens and the observer).
}
\label{fig:figure1}
\end{figure}

\begin{figure}
\epsscale{1.0}
\plotone{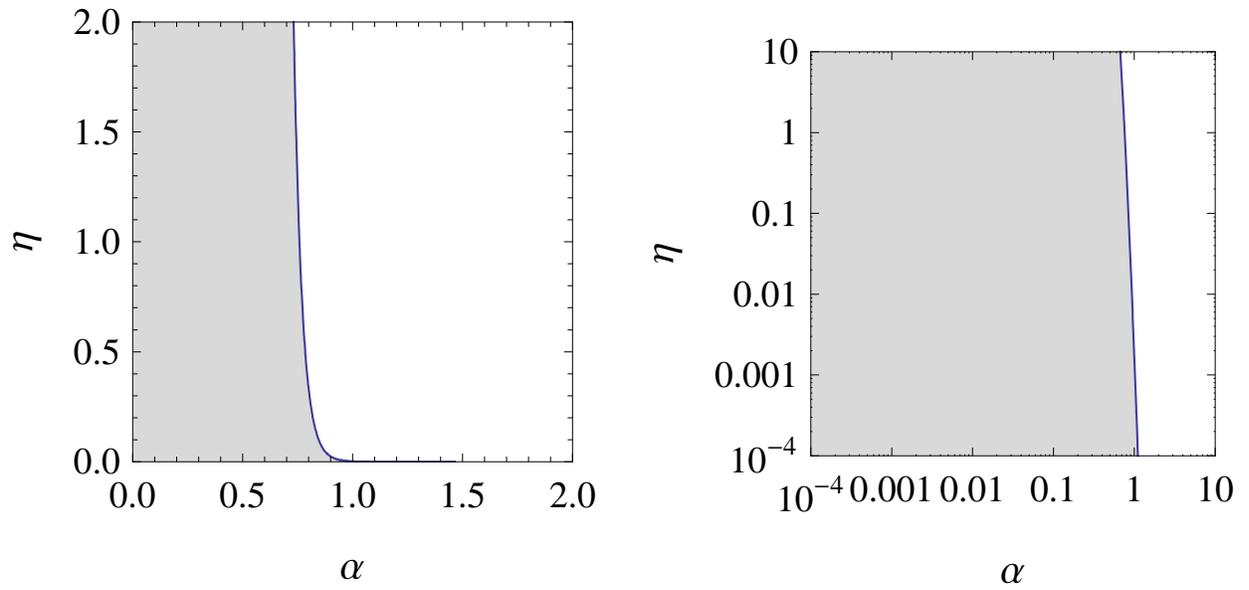}
\caption{
Constraint on the form of the MOND interpolation function
by the atomic interferometer gravitational redshift experiment \citep[][]{Muller2010}.
$\alpha$ and $\eta$ are the parameters in a canonical form of the interpolation function (see Equation~(\ref{eq:canonicalnu})).
The left panel is a linear plot while the right panel is a log-log plot.
The white region is allowed by the experiment and the gray region is excluded.
}
\label{fig:figure2}
\end{figure}

\end{document}